\documentclass[aps,prd,12pt,preprint,tightenlines,unsortedaddress,
amsfonts,amssymb,amsmath,byrevtex,showpacs,nofootinbib]{revtex4-1}
\usepackage{amssymb,amsmath}
\usepackage{latexsym}
\usepackage{graphicx}
\usepackage{geometry}
\usepackage{epstopdf}
\usepackage{color}
\usepackage{epsfig}

\def\wt{\widetilde}
\def\fourthirds{\frac{4}{3}}

\def\ts{\textstyle}
\def\vp{\varphi}
\def\half{\textstyle{\frac{1}{2}}}
\def\quarter{\textstyle{\frac{1}{4}}}

\def\H{{\cal H}}

\def\p{\phi}

\def\H{{\cal H}}

\def\v{\vskip.3cm}

\def\l{\lambda}

\def\t{\textstyle}

\def\ra{\rightarrow}
\def\tint{{\textstyle\int}}

\def\s{\hskip.08em}

\def\d{\partial}

\def\a{\alpha}
\def\b{\begin{eqnarray*}}  
\def\e{\end{eqnarray*}}    
\def\bn{\begin{eqnarray}}  
\def\en{\end{eqnarray}}   

\def\<{\langle}
\def\>{\rangle}
\def\dn{d^n\!x}
\def\no{\nonumber}
\def\ds{d^s\!x}

\def\{{\lbrace}
\def\}{\rbrace}

\begin{document}

\title{Mixed Models: Combining Incompatible Scalar Models in Any Spacetime Dimension}

\author{John R. Klauder}
\email{john.klauder@gmail.com}
\affiliation{Department of Physics and Department of Mathematics, University of Florida,
Gainesville, FL 32611-8440,  USA}

\date{\today}

\begin{abstract}
Traditionally, covariant scalar field theory models are either super renormalizable, strictly renormalizable, or nonrenormalizable. The goal of `Mixed Models' is to make sense of sums of these distinct examples, e.g., $g\vp^4_3+g'\vp^6_3+g''\vp^8_3$, which includes an example of each kind for spacetime dimension $n=3$. We show how the several interactions such mixed models have may be turned on and off in any order without any difficulties.  Analogous results are shown for $g\vp^4_n+g'\vp^{138}_n$, etc.,  for all $n\ge3$. Different categories hold for $n=2$ such as, e.g., ${g P(\vp)_2+g' N\!P}(\vp)_2$, that involve polynomial ($P$) and suitable nonpolynomial ($N\!P$) interactions, etc. Analogous situations for $n=1$ (time alone) offer simple `toy' examples of how such mixed models may be constructed. 

In general, adding a specific classical interaction into a classical free action that reduces the domain of definition leads, effectively, to a
`nonrenormalizable' quantum model. However, the addition of certain classical interactions that, in contrast, do not change the classical domain
of definition, may nevertheless lead to `unsatisfactory' quantum models, which require special treatment. Our discussion will include examples of both
situations.  
\end{abstract}


\maketitle
\section{Toy Models: Classical Story ($n=1$)}
Let us start with elementary examples when $n=1$, namely, in ordinary classical mechanics.
Consider the classical model with the action functional ($g\ge0,\, 0<T<\infty$)
   \bn A_g=\tint_0^T\{\s\half[\s {\dot x}(t)^2- x(t)^2\s]-g\s x(t)^4\s\}\,dt\;, \en
$0<\omega<\infty$, with the domain of $A_g$ given by
   \bn D(A_g)=D(A_0)\equiv\{x(t):\tint({\dot x}^2+x^2 )\s dt<\infty\} \en
   since the given requirement already implies that $\int_0^T x(t)^4\s dt<\infty$ as follows from the
   inequality---where ${x}(t)=(2\pi)^{-1/2}\tint e^{-i\omega t}\s {\wt x}(\omega)\s\s d\omega$---given by
    \bn |x(t)|&&\le (2\pi)^{-1/2}\tint |\s{\wt x}(\omega\s|\,d\omega  \no\\
        && =(2\pi)^{-1/2}\tint(\omega^2+1)^{1/2}\s|{\wt x}(\omega)\s|\,(\omega^2+1)^{-1/2}\,d\omega\no\\
        &&\le(2\pi)^{-1/2}\{\tint(\omega^2+1)\s|{\wt x}(\omega)|^2\,d\omega\cdot\tint(\omega^2+1)^{-1}\s\s d\omega\s\}^{1/2}\no\\
         &&= 2^{-1/2}\s\{\s\tint[\s{\dot x}(t)^2+x(t)^2]\,dt\;\s\}^{1/2}. \en
    Clearly, then, $\lim_{g\ra0}
 A_g=A_0$, both algebraically as well as domain-wise.

 Next consider the classical model with the action functional
 ($g'\ge0$,\, $c>0$)
  \bn A'_{g'}=\tint_0^T\{ \half[\s{\dot x}(t)^2- x(t)^2\s]-g' (x(t)^2-c^2)^{-4}\s\}\,dt\;, \en
  with a domain $D(A'_{g'})=\{x(t):\tint({\dot x}^2+x^2+(x^2-c^2)^{-4}\s)\s dt<\infty\}\equiv D(A'_0)$. In this case, $\lim_{g'\ra0}A'_{g'}=A'_0\not=A_0$. Here, the action functional $A'_0$ is {\it algebraically} the same as $A_0$, but it has a {\it smaller domain} $D(A'_0)\subset D(A_0)$. In short, the model $A'_{g'}$ is {\it not} continuously connected to its own free model $A_0$, but instead it is continuously connected to what we call a {\it pseudofree} model $A'_0$.

  The equations of motion for the pseudofree model are derived from the  algebraic form of the free action functional but their solutions must obey any restricted domain that applies. For the example in the previous paragraph, and for situations with a total energy $E<\half c^2$, the solutions of the free and pseudofree model are identical; only for larger energies are there differences in the solutions of these two models. If $c$ is enormous, e.g., Milky Way sized, then, for all intents and purposes, the solutions of the free and pseudofree models are the same. ({\bf Remark:} Many of the following examples in higher spacetime dimensions will have a similar behavior where there are many---but {\it not} all---identical solutions for their free and pseudofree models.) On the other hand, if $c=0$, then although solutions for the free and pseudofree models with short-time intervals may be the same, {{\it no} long-time solution of the free model is the same as any long-time solution of the pseudofree model.

  Finally, we consider the mixed-model classical action functional ($g\ge0,\,g'\ge0$)
   \bn A''_{g,g'}=\tint_0^T\{\s\half[\s{\dot x}(t)^2- x(t)^2\s]-g x(t)^4-g' (x(t)^2-c^2)^{-4}\s\}\,dt\;,\en
   which includes the two prior interactions. The domain $D(A''_{g,g'})=\{x(t):\tint({\dot x}^2+x^2+(x^2-c^2)^{-4}\s)\s dt<\infty\}\equiv D(A''_0)$. Note that the relevant domain of the mixed model is the {\it smaller} of the domains involved---or the {\it smallest} if more than two interactions are involved. This permits the introduction---and/or removal---of each interaction at any time or order, and leads only to continuous changes whenever any of the interactions is/are involved.

   The quantum story for these models is discussed later.

\section{Abstract Mixed Model Discussion ($n\ge1$)}
 It is straightforward to extrapolate abstractly the main features of the one-dimensional examples to more general classical action functionals that apply to any spacetime dimension $n\ge1$.
The classical action functional for all single-interaction theories of interest has the form $A_\l=A_0+\l\s A_I$, $\l\ge0$, which includes a free action functional ($A_0$), an interaction action functional ($A_I$), and a generic coupling constant ($\l$). The natural assumption is made that $\lim_{\l\ra0}A_\l=A_0$, which algebraically is correct but, perhaps, functionally incorrect. As functionals, each example has a domain $D(A_\l)$ that clearly obeys $D(A_\l)=D(A_0)\cap\s D(A_I)$ for $\l>0$. While sometimes $\lim_{\l\ra0}D(A_\l)=D(A_0)$, it is also possible that $\lim_{\l\ra0}D(A_\l)=D(A'_0)\subset\s D(A_0)$, in which case $\lim_{\l\ra0}\s A_\l=\lim_{\l\ra0}[\s A_0+\l\s A_I\s]=A'_0\not= A_0$. In brief, the interacting action functional $A_\l$ may {\it not} be continuously connected to the free action functional $A_0$ but instead to a pseudofree action functional $A'_0$ with a strictly smaller domain than that of $A_0$. Instead, if one chooses $A'_0$ as the initial unperturbed model, then it follows that $\lim_{\l\ra0}[ A'_0+\l A_I]=A'_0$, namely, the interacting models are continuously connected to the appropriate pseudofree model.

If two interactions are involved at the same time, as in a mixed model, then $A_{\l,\l'}=A_0+\l A_I+\l' A_{I'}$ with a domain $D(A_{\l,\l'})=D(A_0)\cap D(A_I)\cap D(A_{I'})$, and as either of the coupling constants goes to zero, the smallest domain is the relevant one that allows any of the interactions to be continuously introduced, then removed, then introduced again, etc.

As other examples are considered, one may see how this abstract story plays out in all of our examples. We continue the analysis for $n\ge3$, largely because the natural next step ($n=2$) is qualitatively different (and quite interesting as well).

\section{Field Models: Classical Story ($n\ge3$)}
Let us consider an important family of field examples. The general covariant interacting classical scalar model has an action functional given by
   \bn A_g=\tint_0^T\tint \{\half\s[\p_{,\,\mu}(x,t)^2-m_0^2\s\p(x,t)^2\s]-g\s\p(x,t)^p\s\}\,\ds\,dt\;.
   \label{e5}\en
   Here, $p>2$ is any positive even integer, $x\in{\mathbb R}^s$, with the number of spatial dimensions $s\ge2$ and thus a spacetime dimension $n=s+1\ge3$, and
      \bn \p_{,\,\mu}(x,t)^2\equiv {\dot \p}(x,t)^2-[\overrightarrow{\nabla} \p(x,t)]^2 \;,\en
      namely a time derivative squared minus space derivatives squared, as usual. To understand the domain issues, we use a {\it multiplicative inequality} \cite{olga,book} that effectively applies to the Euclidean version of elements of the classical action---thus $x\in{\mathbb R}^n$ here---namely
        \bn \{\tint \p(x)^p\,\dn\s\}^{2/p}\le C_{n,p}\,\tint\{\s[\nabla\p(x)]^2+m_0^2\p(x)^2\}\,\dn\;, \en
        where, for $p\le 2n/(n-2)$ (the renormalizable cases), $C_{n,p}=\fourthirds\s\s m_0^{n(1-2/p)-2}$, while for $p>2n/(n-2)$ (the nonrenormalizable cases), $C_{n,p}=\infty$, meaning in the latter case that there are singular fields for which the integral on the left side diverges while the integral on the right side converges; e.g., $\p_{sing}(x)=|x|^{-\beta}\s e^{-x^2}$, for $n/p\le \beta<(n-2)/2$. As a consequence, for $p\le 2n/(n-2)$, it follows that $D(A_g)=\{\p(x):\tint[(\nabla\p)^2+\p^2]\,\dn<\infty\}\equiv D(A_0)$, while for $p>2n/(n-2)$, it follows that $D(A_g)=\{\p(x):\tint[(\nabla\p)^2+\p^2+\p^p]\,\dn<\infty\}\equiv D(A'_0)\subset D(A_0)$.

        There are many mixed models that can be considered. As one example, consider the classical action functional
          \bn A_{g,g'}=\tint_0^T\tint\{\s\half[\s\p_{,\,\mu}(x)^2-m_0^2\s\p(x)^2\s]-g\p(x)^p-g'\p(x)^{p'}\s\}\,\ds\,dt\;,
          \label{e8}\en
         where, without loss of generality, we assume that $p<p'$. Hence, if $n\ge5$ and $p, p'>2$ are even, positive integers, then it follows that
         \bn D(A_{g,g'})=\{ \p(x):\tint[\s(\nabla\p)^2+\p^2+\p^{p'}\s]\,\dn<\infty\s\}\;. \en

         Observe, even if $g'=0$ initially, but we wish to include the interaction $\p^{p'}$ later,
         then to ensure a smooth, continuous introduction of $\p^{p'}$, we need to `pave the way' for it by already restricting the domain suitably. This anticipation of the term $\p^{p'}$ is equivalent to starting with both interactions and then turning off the interaction $\p^{p'}$ prior to its possible reintroduction.

         In like manner, there are many possible interactions that {\it we will never study}, e.g., $\tint\p(x)^{-4}\s\s \dn, \tint\cos(\p(x))^{-8}\s\s \dn$, etc., and therefore there is absolutely no need to consider the change of domain such never-to-be-introduced interactions  would require.

         The quantum story for these models is discussed later.
\section{Field Models: Classical Story ($n=2$)}}
            The general models discussed in this section are basically the same as in the previous section---e.g., Eqs.~(\ref{e5}) and (\ref{e8})---save for the fact that we focus on spacetime dimension $n=2$, i.e., space dimension $s=1$. We seek to find interactions that are not made finite whenever the Euclidean form of the free action itself is made finite. In brief, we wish to find interactions $V(\p)$ and suitable fields $\p(x)$  such that $\tint[(\nabla\p)(x)^2+\p(x)^2]\,d^2\!x<\infty$ while  $\tint V(\p(x))\,d^2\!x=\infty$.

            With $|x|\equiv (x_0^2+x_1^2)^{1/2}$, fields that have singularities such that $\p'(x)\equiv d\s\p(x)/d|x|\propto |x|^{-1}$ near a singularity
            have the property that $\tint_{|x|<C}\p'(x)^2\s\s |x|$ $ d|x|\propto\tint_{|x|<C}\s |x|^{-1}\s\s d|x|=\infty$, for $0<C<\infty$; thus fields of the form $\p(x)\simeq \ln(|x|)$, $|x|\le C$, lead to free action functionals that are divergent due to the derivative term. Thus, to ensure finite free action functionals, we need to have behavior such as $\p'(x)\simeq |x|^{-1}\s[ \ln(|x|)]^{-\delta}$ for $\delta>\half$ since then $\tint_{|x|<C}\p'(x)^2\s|x|\s\s d|x|\simeq\tint_{|x|<C}|x|^{-1}\s[\ln(|x|)]^{-2\delta}\s\s d|x|=\tint^\infty z^{-2\delta}\s\s dz<\infty$, where $x=e^{-z}$. This means that ``good'' fields with finite free actions include $\p(x)\simeq [\ln(|x|)]^{1-\delta}$. For such ``good'' fields to lead to divergent interaction action functionals, we suggest that $V(\p(x))=\exp[ \beta\p(x)^2]$, where $\beta>2$.

            We observe that fields of the form $\p(x)=\p_\epsilon(x)\equiv (\s|x|^{\epsilon}-1\s)/\epsilon$, $\epsilon>0$, for a region where $|x|<C$, and which smoothly fall to zero for $|x|>C$ in a compact region, have finite integrals for the free action terms. This property follows from the fact that
              \bn \tint_{|x|<C}\, \p'_\epsilon(x)^2\,|x|\, d|x|=\tint_{|x|<C} \,|x|^{-1+2\epsilon}\,d|x|<\infty\;.\en
              Possible interaction of the form $V(x)=\p_\epsilon(x)^p$ all satisfy $\tint_{|x|<C}\,\p_\epsilon(x)^p\,d|x|<\infty$. On the other hand, $V(x)=e^{\beta\s\p_\epsilon(x)}$ leads to
              \bn \tint_{|x|<C}\,e^{[\beta[|x|^\epsilon-1\s]/\epsilon)}\;d|x|<\infty\;.\en
              However, if one is interested in integration over the entire space, it is important to replace potentials of the form $e^{\alpha\s\phi(x)}$ by
              the form $e^{\alpha\s\phi(x)}-1$, as well as related potentials.

\section{Toy Models: Quantum Story ($n=1$)}
          The quantum theory of the classical model
          \bn A_g=\tint_0^T\{\s\half[{\dot x}(t)^2- x(t)^2\s]-g\s x(t)^4\s\}\,dt \en
          is well described by the Schr\"odinger equation
            \bn i\hbar\s\d\s\psi(x,t)/\d t=-\half\hbar^2\s\d^2\s\psi(x,t)/\d x^2+\half x^2\s\psi(x,t)+g\s x^4\s\psi(x,t)\;,\en
            which exhibits solutions that are continuous in $g\ge0$ implying that the interacting quantum theory is continuously connected to the quantum free theory. This property may be stated differently, namely as
            \bn \<x'',T|x',0\>_0 &&=\lim_{g\ra0}{\cal N}\int_{x(0)=x'}^{x(T)=x''}\,e^{\t (i/\hbar)\tint_0^T\{\half[{\dot x}^2-
             x^2]-g\s x^4\}\,dt}\;{\cal D}x\no\\
              &&={\t\sum}_{m=0}^\infty h_m(x'')\,h_m(x')\,e^{\t-(i/\hbar)[\hbar(m+1/2)T]}\;,\en
              where $\{h_m(x)\}$ denote conventional Hermite functions. This relation demonstrates
              that the quantum propagator reduces to that of the free theory as $g\ra0$.

              On the other hand, the quantum theory of the classical model
              \bn A'_{g'}=\tint_0^T\{\s\half[\s{\dot x}(t)^2- x(t)^2\s]-g'[x(t)^2-c^2]^{-4}\s\}\,dt \;,\en
              with $c>0$, is well described by the Schr\"odinger equation
              \bn &&i\hbar\s\d\s\psi(x,t)/\d t=-\half\hbar^2\s\d^2\s\psi(x,t)/\d x^2_{DBC\;(-c,\s c)}+\half x^2\s\psi(x,t)\no\\
              &&\hskip12em +g'\s (x^2-c^2)^{-4}\s\psi(x,t)\;,\en
              where $DBC\,(-c,c)$ denotes Dirichlet boundary conditions, i.e., $\psi(x,t)=0$, at $x=\pm\s\s c$ \cite{simon}. Such boundary conditions arise automatically due to the singularity of the potential at $x=\pm\s\s c$, and {\it these boundary conditions survive when $g'\ra0$, leading to the behavior of the quantum pseudofree model}. This means that the propagator for the pseudofree model is {\it not} the propagator for the free model, i.e.,
               \bn \<x'',T|x',0\>'_0\s     
               &&=\lim_{g'\ra0}{\cal N}\int_{x(0)=x'}^{x(T)=x''}\,e^{\t (i/\hbar)\tint_0^T\{\half[{\dot x}^2- x^2]-g'(x^2-c^2)^{-4}\}\,dt}\;{\cal D}x\no\label{er5}\\
                  &&\not={\t\sum}_{m=0}^\infty h_m(x'')\s h_m(x')\,e^{\t-(i/\hbar)[\hbar(m+1/2)T]}\;. \en
                  When $c>0$, as we have assumed, the explicit functional form of the pseudofree propagator generally involves unknown functions and unknown energy levels. However, if $c=0$, so that the Dirichlet boundary conditions apply only to $x=0$, it follows that
                  \bn \<x'',T|x',0\>'_0\s &&=\lim_{g'\ra0}{\cal N}\int_{x(0)=x'}^{x(T)=x''}\,e^{\t (i/\hbar)\tint_0^T\{\half[{\dot x}^2- x^2]-g'(x^2)^{-4}\}\,dt}\;{\cal D}x\no\\
                  &&={\t\sum}_{m=0}^\infty h_m(x'')\s h_m(x')\,[1-(-1)^m]\,e^{\t-(i/\hbar)[\hbar(m+1/2)T]}\;, \no\\ \en
                  which, although the pseudofree propagator involves some of the same (i.e., the odd) Hermite functions and energy levels, it clearly differs from the free propagator.

                  From a classical point of view, the singular potentials in these last two examples lead to a different domain of the pseudofree theory and the free theory. In a similar manner, and now
                  viewed from a path integral point of view, the same singular potentials in these last two examples act partially as {\it hard cores} projecting out certain paths that are not recovered when $g'\ra0$, leading thereby to the pseudofree behavior.

                  As a quantum mixed model we start with the classical model
                    \bn A_{g,g'}=\tint_0^T\{\half[{\dot x}(t)^2-x(t)^2]-g x(t)^4-g'[x(t)^2-c^2]^{-4}\s\}\,dt\;,\en
                    assuming that $c>0$, which leads to the Schr\"odinger equation
                    \bn &&i\hbar\s\d\s\psi(x,t)/\d t=-\half\hbar^2\d^2\s\psi(x,t)/\d x^2_{DBC\,(-c,\,c)}+\half x^2\s\psi(x,t)\no\\
                    &&\hskip8
                    em +gx^4\psi(x,t)+g'(x^2-c^2)^{-4}\psi(x,t)\;, \en
                    where the boundary conditions only refer to the latter interaction. Consequently, the pseudofree model for this example is the same as (\ref{er5}), exactly as if the first interaction had been absent altogether.

\section{Field Examples: Quantum Story ($n\ge3$)}
      As our first set of examples we consider classical models with actions given by
      \bn A_g=\tint_0^T\tint\{\s\half[\s\p_{,\,\mu}(x,t)^2-m_0^2\p(x,t)^2\s]-g\s\p(x,t)^p\s\}\,\ds\,dt\;,\en
      where $n\ge3$ ($s\ge2$) and $p>2n/(n-2)$, which refers to the {\it nonrenormalizable models}. According to the multiplicative inequality discussed in Sec.~3, such models have pseudofree classical theories that differ from their own free classical theories. Consequently, these interacting classical models are {\it not} continuously connected to their own free classical theory. If this fact holds classically, there is every reason to believe it holds for the quantum theory as well, namely, the zero coupling limit of the interacting quantum theory is not continuously connected to the quantum free model, just as we saw happen for the toy models. On the other hand, standard perturbation procedures assume that the interacting quantum theory {\it is} continuously connected to the free quantum theory; if, as we claim, this assumption is false, then divergences in a traditional perturbation analysis may well be expected.

      Elsewhere we have argued that if one assumes that nonrenormalizable models have a quantum pseudofree theory which differs from the usual quantum free theory, then the associated interacting models:~(i) may be well defined, (ii) may be nontrivial, and (iii) may have the proper, nontrivial classical limit. We need to choose a counterterm, which leads to a suitable pseudofree theory that, besides continuity of the zero
      coupling-constant limit of the interacting models, has the desirable property that a power-series expansion about the pseudofree theory is divergence free, and, moreover, the classical limit of an interacting quantum model yields the expected nonlinear classical model. In this paper we content ourselves with describing such a proper counterterm and listing a few of its properties; how such a proper counterterm was actually discovered has already been discussed elsewhere, see, e.g., \cite{slavnov, newbook}, and related references therein.

      Based on using the special counterterm for such models, the lattice regularized Hamiltonian operator for interacting models is given by
      \bn \H\, &&=-\half\hbar^2\s a^{-2s}{\t\sum}'_k\,\frac{\d^2}{\d\p_k^2}\,a^s+\half{\t\sum}'_{k,k^*}(\p_{k^*}-\p_k)^2\s a^{s-2} \no\\
    &&\hskip2em+\half\s m_0^2{\t\sum}'_k\p_k^2\,a^s+
    g_0{\t\sum}'_k\p_k^p\,a^s
    +\half\hbar^2{\t\sum}'_k{\cal F}_k(\p)\,a^s-E_0\;. \label{eH}\en
    In this expression, the counterterm is proportional to $\hbar^2$, and specifically is chosen so that
    \bn {\cal F}_k(\p)\, && \equiv\quarter\s(1-2ba^s)^2\s
          a^{-2s}\s\bigg({{\ts\sum}'_{\s t}}\s\frac{\t
  J_{t,\s k}\s \p_k}{\t[\Sigma'_m\s
  J_{t,\s m}\s\p_m^2]}\bigg)^2\no\\
  &&\hskip2em-\half\s(1-2ba^s)
  \s a^{-2s}\s{{\ts\sum}'_{\s t}}\s\frac{\t J_{t,\s k}}{\t[\Sigma'_m\s
  J_{t,\s m}\s\p^2_m]} \no\\
  &&\hskip2em+(1-2ba^s)
  \s a^{-2s}\s{{\ts\sum}'_{\s t}}\s\frac{\t J_{t,\s k}^2\s\p_k^2}{\t[\Sigma'_m\s
  J_{t,\s m}\s\p^2_m]^2}\;. \label{eF} \en
  Here $b$ is a fixed, positive parameter with dimensions (Length)$^{-s}$ and $J_{k,l}\equiv 1/(2s+1)$ for $l=k$ and for $l$ each one of the $2s$ spatially nearest neighbors to $k$; otherwise, $J_{k,l}\equiv 0$.
  Although ${\cal F}_k(\p)$ does not depend only on $\p_k$, it nevertheless becomes a local potential
  in the formal continuum limit.

  It should be clear that the chosen counterterm 
  is completely different from the usual counterterm(s) of conventional canonical quantization. In fact, the counterterm (\ref{eF}) has been chosen to {\it remove the source of all divergences} rather than cancel divergences, one after another,  as they arise in a perturbation series analysis. For a brief discussion of how  the counterterm (\ref{eF}) was chosen, see \cite{slavnov}.

  The lattice-regularized action functional for such models follows directly from the form of the Hamiltonian operator, and is given by
    \bn &&I=\half \Sigma_{k,k^*}(\p_{k^*}-\p_k)^2\,a^{n-2}+\half m_0^2\Sigma_k\p_k^2\s a^n\no\\
     &&\hskip5em +g_0\Sigma_k\p_k^4
     \s\s a^n+\half\Sigma_k \hbar^2{\cal F}_k(\p)\s\s a^n\;, \label{eL}   \en
     where the sum is over the $n$-dimensional lattice and $k^*$ denotes one of the $n$ nearest neighbors to $k$ in the positive sense.
     This expression is suitable to use in lattice-regularized, Euclidean functional-integral expressions.

     Regarding the general choice of counterterms, recall that canonical quantization is generally open to ${\cal O}(\hbar)$ terms as part of the Hamiltonian operator. More directly, the program of enhanced quantization \cite{enh,newbook} specifically allows quantization procedures that permit unconventional counterterms such as (\ref{eF}).

     Apart from a possible set of measure zero, it is important to observe that the models under consideration that are continuous perturbations of the free model for the classical story are also continuous perturbations of the free model for the quantum story. Likewise, those models that are discontinuous perturbations of the free model for the classical story are also discontinuous perturbations of the free model for the quantum story. In each case, this behavior may be understood within a functional integral approach to quantization as an interaction term that either does not or does partially act as a hard core projecting out a set of functions that has a positive measure. This argument is similar to the reason that singular potentials led to pseudofree behavior in the `toy' models of Secs.~1 and 5. \v

     {\bf [Remark:} There are valid arguments \cite{aiz,fro} which show that nonrenormalizable quantum models $\vp^4_n$, $n\ge5$, are {\it trivial} ($=$ free) models based on the continuum limit of appropriate lattice functional integrals, and thus the classical limit of the quantum model is also free, which conflicts with the original nontrivial classical model. These arguments are based on standard procedures of allowing only counterterms for mass, field strength, and coupling constant, and implicitly assume
     the interacting models are continuously connected to their own free model. On the other hand, the interacting lattice action functional (\ref{eL}) contains an unconventional counterterm and it does not pass to a free lattice action functional as the coupling constant $g_0$ vanishes. The continuum limit of a functional integral using the lattice action  (\ref{eL}) is difficult to treat analytically, but, hopefully, Monte Carlo studies can shed some light on its properties.
     In that regard, preliminary, but limited, Monte Carlo results \cite{stan} do suggest nontriviality.{\bf]} \v

     It is important to observe that the chosen counterterm (\ref{eF}) does not depend on the coupling constant $g_0$ nor on the interaction term power $p$; however, it does depend on the spatial dimension $s$ and thus on the spacetime dimension $n=s+1$. So far our discussion has focussed on nonrenormalizable models, but, just as was the case for the toy models, we can extend the use of the new counterterm to renormalizable models as well for which $p\le2n/(n-2)$. {\bf{[Remark}}: The conventional solution for any (super)renormalizable model still applies to suitable physical applications where nonrenormalizable interactions would never arise.{\bf]} Thus, for a given value of $n=s+1$, this counterterm can be used for {\it all} interactions $\vp^p_n$, where $p$ is generally a positive, even number, i.e., both renormalizable and nonrenormalizable models, and thus the special counterterm (\ref{eF}) applies for mixed models such as $g\vp_n^p+g'\vp_n^{p'}+g''\vp_n^{p''}$, etc., so long as the sum is suitably bounded below.

     Note well that there do exist further models with interactions such as $\p(x)^{-4}$, etc., that we {\it{explicitly forbid}} classically, and therefore we do not need to 
     accommodate such models in the quantum story.

\section{Field Examples: Quantum Story ($n=2$)}
In Sec.~4, the classical story for $n=2$ fields showed that interactions of the form $g\exp[\a\p(x)]$, $0\le g$, and for all $\a$, $0\le\a<\infty$, were {\it continuous perturbations of the free theory} in the sense that as $g\ra0$ the action functional for the interacting model
\bn   A_g(\p)=\tint_0^T\tint\s\{\s \half[\s\p_{,\s\mu}(x,t)^2-m_0^2\,\p(x,t)^2\,]-g\s[\s e^{\a\s\p(x,t)}-1\s]\,\}\,dx\,dt \en
converges to the action functional for the free models $A_0(\p)$ and importantly, the domain of the interacting action functional $D_g(\phi)$ is equal to the domain of the free action functional $D_0(\p)$, i.e, now with $x\in \mathbb{R}^2$,
\bn    &&D_g(\p)=\{\p(x)\s:\s\tint[\s\nabla\p(x)^2+\p(x)^2+g\s(\s e^{\a\p(x)}-1\s)\s] \,d^2x<\infty\}\no\\
        &&\hskip2.92em=\{\p(x)\s:\s\tint[\s\nabla\p(x)^2+\p(x)^2\s] \,d^2x<\infty\}=D_0(\p)\;.\en

Unlike other examples discussed earlier in this paper, the continuity properties of the classical theory are different than the continuity properties of the conventional quantum theory. For example, it is known \cite{sergio} that for quantum interactions of the form $g\s:\exp[\a\s\hat{\p}(x)]:$, the interacting quantum theory is  continuously connected to the free quantum theory only for $\a^2<4\pi$. On the other hand, for $\a^2\ge 4\pi$, and $g>0$, the interaction leads to a {\it free} theory \cite{sergio}, not unlike the result for $\p^4_n$ and $n\ge5$ when dealing with a lattice limit of the model \cite{aiz,fro}. This situation means that while the classical theory behaves as expected that is not the case for the quantum theory. Thus, for the models in question when $n=2$, we see that we have a very different relationship between the quantum and classical behavior of covariant scalar models than was the case previously.

{\bf{[Remark:}} This kind of behavior---let us call it ``good classical, bad quantum''---is known to arise in some other (non-covariant) models; for example, models with the classical Hamiltonian $H(p,q)=\half\Sigma_l[\s p_l^2+m_0^2\s q_l^2]+ g\s(\Sigma_lq_l^2\s)^2$, where $1\le l\le\infty$ (a so-called rotationally symmetric model). Another class of models with a similar behavior has the classical Hamiltonian $H(p,q)=\half[\s {\rm{Tr}}(p^2) + m_0^2\s{\rm{Tr}}(q^2)\s]+ g\s {\rm{Tr}}(q^4)$, where in this case $p=\{p_{mn}\}$ and $q=\{q_{mn}\}$, where $p_{nm}=p_{mn}$ and $q_{nm}=q_{mn}$, when $1\le m,n\le\infty$ (a so-called rotationally symmetric tensor model). When either of these models is quantized in the standard way, the result is that of a {\it free model} whenever an infinite number of variables is involved. On the other hand, it is obvious that the domain of the interacting classical Hamiltonian is the same as the domain of the corresponding free classical Hamiltonian; such classical models encounter no difficulties for an infinite number of variables whatever the value of $g>0$, and they lead to genuinely interacting, non-free, classical models. How triviality of the conventional quantum procedures arises and how nontriviality of the  unconventional quantum procedures arises in these two cases is spelled out in \cite{klaRS,newbook}{\bf]}.

\section{Conclusion}
We have shown that certain classical interacting models become their own free models when the interaction term is reduced to zero, while other classical interacting models become pseudofree models with strictly smaller domains of definition than that of the free model when the interaction term is reduced to zero. Interacting classical models that become pseudofree classical models in such a limit lead to interacting quantum models that become pseudofree quantum models differing from their own free quantum models. Quantum models that are nonrenormalizable when treated conventionally exhibit divergences because they are assumed to possess an expansion about the free theory when in fact their zero coupling limit is a pseudofree model about which a perturbation exists without divergences. For models with several different interactions, all for the same number of spacetime dimensions, it is found that a suitable pseudofree model exists for which the several interactions can be turned on and off in any order without any unusual behavior. In most cases this acceptable quantum behavior has its footprint already in its classical behavior. However, there are certain cases, most notable when the spacetime dimension $n=2$, for which a pseudofree quantum behavior does not automatically imply a pseudofree classical behavior but rather leads to a free classical behavior when its interactions are removed.

\section*{Acknowledgements}
The author thanks Sergio Albeverio and Ludwig Faddeev for helpful correspondence.

\end{document}